\documentclass[preprint,aps,preprintnumbers,amsmath,amssymb]{revtex4}
\usepackage{graphicx}
\begin{document}
\draft
\title {Thermodynamics of an incommensurate quantum crystal}
\author {P. W. Anderson, W. F. Brinkman and David A. Huse}
\address{Department of Physics, Princeton
University, Princeton, NJ 08544}
\begin{abstract}
We present a simple theory of the thermodynamics of an
incommensurate quantum solid.  The ground state of the solid is
assumed to be an incommensurate crystal, with quantum zero-point
vacancies and interstitials and thus a non-integer number of atoms
per unit cell.   We show that the low temperature variation of the
net vacancy concentration should be as $T^4$, and that the first
correction to the specific heat due to this varies as $T^7$; these
are quite consistent with experiments on solid $^4$He.  We also make
some observations about the recent experimental reports of
``supersolidity'' in solid $^4$He that motivate a renewed interest
in quantum crystals.
\end{abstract}
 \maketitle

Recent experiments 
\cite{kimchan1, kimchan2} showing a marked low temperature reduction
in the rotational moment of inertia of crystals of solid $^4$He have
rekindled interest in this highly quantum-mechanical solid.  The
proposed ``supersolid'' phase is believed to occur due to the
quantum behavior of point defects, namely vacancies and
interstitials, in this crystal of bosons \cite{andreev}.  Over the
past twenty years various experiments have been performed to measure
the temperature dependence of the vacancy concentration in solid
$^4$He.  Although there are considerable differences in the results,
the most accurate data comes from x-ray measurements of the lattice
constant as a function of temperature at fixed density
\cite{fraass4}.  These data have usually been interpreted in terms
of a classical theory of vacancies involving an activation energy
and a configurational entropy for their creation.  However, this
theory implies a corresponding vacancy contribution to the specific
heat that is as large as the phonon contribution near 1 Kelvin
\cite{burns}.  Such a classical vacancy contribution to the specific
heat has not been seen; the specific heat is instead well explained
almost entirely in terms of the $T^3$ term from the phonon spectrum,
and the leading correction to this fits very well to a $T^7$ term
\cite{heat4}. There have been various attempts to explain this
discrepancy but none have been satisfactory and the problem has
remained open \cite{burns}.

Here we propose a simple phenomenological thermodynamic description
of a low-temperature incommensurate quantum solid.  We note that the
ground state of a quantum solid need not be commensurate, i.e. it
need not have an integer number of atoms per unit cell
\cite{andreev}.  One description of the quantum solid is 
as a density wave that has formed in the quantum fluid. The
periodicity of this density wave need not match precisely to the
particle density, so that the ground state may be incommensurate,
with unequal densities of vacancies and interstitials.  The x-ray
measurements on solid hcp $^4$He show that the density of vacancies
increases faster than that of interstitials with increasing
temperature \cite{fraass4}, indicating that thermal fluctuations
favor vacancies more than interstitials. Whether or not the same is
true for quantum fluctuations is not clear at this point.  We
develop a simple thermodynamic theory of the low temperature
behavior of an incommensurate quantum solid, finding that the low
temperature net change in vacancy density at fixed particle density
follows a $T^4$ power law behavior.  The x-ray data are quite
consistent with such a temperature dependence, as we show below.  In
addition, we show that this simple model produces a $T^7$ correction
to the specific heat, as has been observed \cite{heat4}. Such a
scenario could apply to any highly quantum solid and is not specific
to bosons, so solid $^3$He should and does show similar phenomena
\cite{vac3, heat3}; perhaps hydrogen might, also.

It has been argued by one of us based on Jastrow-type wave functions
that it is expected that there will be vacancies in the ground state
of a highly quantum fluctuating solid such as $^4$He, and that such
a ground state may be superfluid \cite{pwa}.  The vacancies are an
integral feature of the ground state and carry no entropy or energy.
These vacancies may be sufficiently mobile that they never behave as
classical particle-like objects at the temperatures where the solid
is present.  Thus we will assume that the vacancies and
interstitials in solid $^4$He remain in a strongly-correlated
quantum state up to temperatures in the vicinity of 1 Kelvin, so
they do not make a large contribution to the specific heat other
than the incommensurability effect that we describe below.

For an hcp lattice of volume $V$ and lattice constant $a$, the
number of lattice sites is $N_s=V\sqrt{2}/a^3$.  If the ground state
crystal is incommensurate, then its number $N$ of atoms differs from
its number of sites: $N\neq N_s$.  Recent data on the possible
superfluid nature of these solids \cite{kimchan2} suggests that
these two numbers could differ by up to 1\%, although it seems quite
possible that the 1\% effect in the apparent superfluid density
could arise from a much smaller (or even zero) net density of
defects. Such a small difference between the number of atoms and the
number of lattice sites may have escaped detection in simulations
\cite{sim1, sim2, sim3} of the ground state of solid $^4$He.  Direct
comparisons of experimental measurements of the density of $^4$He
atoms to the x-ray density of sites do not appear to have been
published for the low pressure hcp phase where the apparent
supersolidity has been seen, although Simmons \cite{simmons} tells
us that the difference appears to be well under 1\%. We thus
strongly urge that more simultaneously precise density and lattice
constant measurements be done for the quantum solids to learn how
incommensurate their ground states really are, especially at the
lowest densities where quantum fluctuations should be strongest.

Given that a crystal may be incommensurate, one needs to develop a
theory in which the lattice constant and the density can change
independently.  In the temperature range we consider, the vacancies
and interstitials are assumed to be incorporated in a
highly-correlated quantum state of the system and the only low
frequency modes giving large contributions to the temperature
dependence of the free energy are the phonons.  In the standard
treatment of the low temperature thermal expansion of a crystal it
is the density dependence of the phonon velocities (the Gruneisen
parameters) that determine the expansion.  Here we will instead work
at fixed particle density, but allow the lattice constant and thus
the incommensurability to vary, driven by the dependence of the
phonon velocities on the incommensurability.  Thus we consider the
free energy for a given mass of helium at a fixed volume, so that we
do not need to include the overall density as a variable.  Let the
incommensurability
\begin{equation}
\epsilon=\frac{N_s-N}{N_s}=\epsilon_0+\delta
\end{equation}
be the net fractional vacancy number (i.e., the fraction of
vacancies minus the fraction of interstitials).  We will ask about
the crystal's behavior as a function of its incommensurability,
although this is not a variable that is under ready experimental
control.  Here $\epsilon_0$ is the incommensurability at absolute
zero temperature. Thus we obtain the following expression for the
free energy as an expansion at low temperature and low deviation
$\delta$ of the incommensurability from the ground state value:
\begin{equation}
F=-E_0+\frac{E_2}{2}\delta^2-(D_0+D_1\delta+...)T^4+...~.
\end{equation}
$-E_0$ is the ground state energy and $E_2$ gives the harmonic
increase of the crystal's energy when, staying at $T=0$, the
incommensurability is changed away from its ground state value by
changing the number of lattice sites and thus the lattice constant.
The $T^4$ term in the free energy is simply that due to acoustic
phonons, and possibly also the acoustic superfluid mode that is
expected to be present in a supersolid.  The velocities of these
acoustic modes in general vary with the incommensurability and are
not at an extremum at the ground state incommensurability (which is
$\delta=0$).  Thus there is a term that is linear in $\delta$ in the
prefactor of this $T^4$ term, from its lowest-order linear variation
with the incommensurability $\epsilon$.  The parameter $D_1$ plays
the role of the Gruneisen parameter in driving the change in the
lattice constant with temperature, but here this change is happening
at fixed particle density.  Next we simply find the value of the
incommensurability that minimizes this free energy (2) at a given
low temperature, obtaining to lowest order the temperature
dependence
\begin{equation}
\delta\approx\frac{D_1}{E_2}T^4~,
\end{equation}
instead of the classical thermally-activated form
($\delta\sim\exp{(-\Delta/k_BT)}$) that one obtains in the classical
vacancy theory.  Figure 1 shows that the x-ray data 
\cite{fraass4} fits about as well to our proposed $T^4$ temperature
dependence as it does to the classical theory.  
Clearly, when similar measurements are made more precisely and/or
carried to lower temperatures, a discrimination between these two
simple theories will be made; again, we encourage such efforts. Note
that here $\delta$ is the increase in the fractional net density of
vacancies above the possibly nonzero value it already has in the
ground state.  Also, the strong quantum fluctuations might mean that
the ground state concentrations of vacancies and interstitials are
both rather larger than $\epsilon_0$, but it is only the difference
between these densities (that we are calling the net density) that
is readily measurable and that enters as a thermodynamic parameter.

\begin{figure}
\includegraphics[width=3.25in]{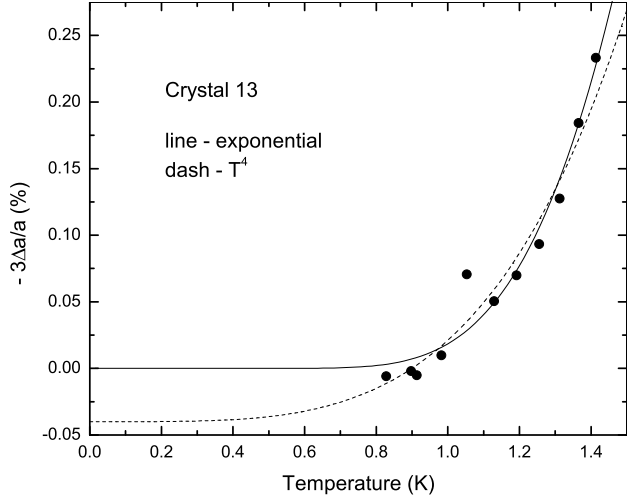}
\caption{\label{fig} The percentage net density of vacancies in a
solid $^4$He crystal of molar volume 20.9 cm$^3$, as measured using
x-rays via the change $\Delta a$ in the lattice constant from a
reference value. Filled circles are the data from
Ref. \cite{fraass4}. The solid line is a 
thermally-activated (classical) fit, 
while the dashed line is a 
fit to the $a\approx a_0+bT^4$ behavior we expect if the ground
state is incommensurate. The zero on the vertical axis is free and
was chosen so that the classical fit goes to that value in the low
$T$ limit. Figure courtesy of Ralph Simmons. }
\end{figure}

If instead the ground state is commensurate ($\epsilon_0=0$) and is
locked in to a Mott ``insulating'' state with exactly one atom per
lattice site, then the energy as a function of the change in the
incommensurability $\delta$ is linear ($\sim |\delta|$) rather than
quadratic.  This results in the classical, thermally-activated
behavior (it can be viewed as activation of atomic ``carriers''
across the Mott gap of this insulator).  Another possibility is that
the quantum fluctuations are strong enough to put the system out of
the Mott insulating phase, but the ground state remains commensurate
with $\epsilon_0=0$ due to an approximate (one might say
``coincidental'') vacancy-interstitial symmetry of the ground state.
In this latter case, the $\delta\sim T^4$ behavior will occur,
provided the thermal excitations break that approximate symmetry, as
they certainly appear to from the x-ray data (Fig. 1)
\cite{fraass4}.

A second known anomaly follows from (2).  The specific heat of solid
$^4$He in the temperature range near 1 Kelvin was shown 
to fit nearly exactly (see \cite{heat4}, Fig. 7) to the sum of two
power laws:
\begin{equation}
C=AT^3+BT^7~.
\end{equation}
The phonons give corrections to the $T^3$ specific heat due to their
anharmonicity and dispersion, but these are expected to be down from
the leading $T^3$ by powers of $(T/\Theta_D)$, where the Debye
temperature $\Theta_D\cong 25$ K for helium.  The observed $T^7$
correction is orders of magnitude larger than this \cite{heat4}.
Minimizing (2) with respect to $\delta$, the free energy as a
function of temperature behaves as
\begin{equation}
F=-E_0-D_0T^4-D_1^2T^8/2E_2+...~.
\end{equation}
Thus the incommensurate crystal shows a positive $T^7$ leading
correction to the phonon specific heat, due to its change of
incommensurability with temperature.  This is quite consistent with
the experimental specific heat measurement \cite{heat4}.  The x-ray
and specific heat experiments together give rough estimates of
$E_2\cong 80$ K/atom, $D_0\cong 0.013$ (K$^3$-atom)$^{-1}$ and
$D_1\cong 0.06$ (K$^3$-atom)$^{-1}$ for the parameters in our free
energy (taking $k_B=1$).  The new parameters $E_2$ and $D_1$ are of
the same order as but larger than $E_0$ and $D_0$, respectively, all
of which seems quite reasonable to us.

It should be noted that nowhere in the present argument did we
invoke the boson nature of $^4$He.  In fact, the discrepancies found
in $^4$He between the temperature dependent x-ray vacancy data and
the specific heat data within a classical vacancy model are also
there in solid $^3$He \cite{vac3}.  There are quantitative
differences between the isotopes, however, in that the corrections
to the leading $T^3$ in the specific heat are much larger in $^3$He
\cite{heat3}.  In fact, for $^3$He the correction to the leading
$T^3$ term becomes larger than the $T^3$ term itself, and does not
fit well to a simple $T^7$ correction \cite{heat3}.  But when the
correction is that large, it should be expected that terms beyond
$T^7$ cannot be neglected.  Of course, the difference between bosons
and fermions is essential when considering supersolidity, but it is
not crucial for the thermodynamic issues we have discussed above.

Before concluding, we make a few comments about the recent
experimental indications \cite{kimchan2} of ``supersolid'' behavior
in solid $^4$He:  First we note the strong dissipation feature seen
in the amplitude of their oscillator vs. temperature in Fig. 2A of
Ref. \cite{kimchan2}. This dissipation should be significant only
when the rate of damping of the superflow is of the same order as
the frequency of the oscillator, which is about 1 kHz.  The broad
(on the temperature axis) dissipation feature implies that this
damping rate is decreasing rather gradually with decreasing
temperature, and passes through 1 kHz near the maximum damping,
around $T=60$ mK. The appearance of a detectable apparently
supersolid signal at much higher temperature should not be viewed as
a possible supersolid phase transition at those higher temperatures,
but instead possibly as the temperature where the precursors to
supersolidity (the critical fluctuations) first become detectable in
this experiment.  This very broad regime with precursors
to the apparent supersolidity 
suggests to us two possibilities: first, that perhaps these
experiments are near a supersolid quantum critical point, where the
quantum fluctuations destroy supersolid order in the ground state,
replacing it with some sort of quantum vortex liquid ground state;
or second, that the superflow is being damped by some
temperature-dependent mechanism other than vortices (transverse
phonons and {\it umklapp} are two possibilities that are not present
in the liquid phase) and this damping only vanishes at zero
temperature.  Note that here we are always discussing the damping at
linear order in the apparent superfluid velocity, thus in linear
response to the solid's motion.  The actual supersolid transition is
where this rate of damping vanishes, so one can have a true
superflow in linear response. From these recent experiments 
at just the one frequency \cite{kimchan2}, we cannot determine where
this transition actually happens, or even whether it does happen
even at zero temperature, although we should conclude from their
data that the supersolid transition temperature must be below the
dissipation feature, which puts it below 50 mK \cite{breppy}. The
results of similar experiments at other frequencies that are ``in
the works'' should be very informative.  \cite{ack}

\end{document}